\documentclass[onecolumn]{elsart3-1}

\usepackage{indentfirst}
\usepackage[english,francais]{babel}
\usepackage[latin1]{inputenc}

\usepackage{graphicx}

\usepackage{amsfonts}
\usepackage{amsmath}
\usepackage{amssymb}

\renewcommand{\theequation}{\arabic{equation}}
\def\og{\leavevmode\raise.3ex\hbox{$\scriptscriptstyle\langle\!\langle$~}}
\def\fg{\leavevmode\raise.3ex\hbox{~$\!\scriptscriptstyle\,\rangle\!\rangle$}}

\newcommand{\pd}[2]{\frac{\partial#1}{\partial#2}}
\newcommand{\abs}[1]{\left|#1\right|}

\def\k{\vec{k}}
\def\g{\vec{g}}
\def\v{\vec{v}}
\def\u{\vec{u}}
\def\q{\vec{q}}

\def\0{\vec{0}}
\def\n{\vec{n}}
\def\x{\vec{x}}
\def\C{\vec{C}}
\def\fphi{\hat{\phi}}
\def\feta{\hat{\eta}}
\def\vpsi{\vec{\psi}}
\def\fvpsi{\hat{\vpsi}}
\def\fpsi{\hat{\psi}}
\def\R{\mathbb{R}}
\def\L{\mathcal{L}}
\def\O{\mathcal{O}}
\def\F{\mathcal{F}}
\def\grad{\nabla}
\def\div{\nabla\cdot}
\def\divx{\nabla_{\x}\cdot}

\def\ssigma{\underline{\underline{\vec{\sigma}}}}

\begin{document}

\begin{frontmatter}

\selectlanguage{english}

\title{Viscous potential free-surface flows in a fluid layer of finite depth}
\author{Denys Dutykh},
\ead{Denys.Dutykh@cmla.ens-cachan.fr}
\ead[url]{www.cmla.ens-cachan.fr/\~{}dutykh}
\author{Frédéric Dias}
\ead{dias@cmla.ens-cachan.fr}
\ead[url]{www.cmla.ens-cachan.fr/\~{}dias}
\address{CMLA, ENS Cachan, CNRS, PRES UniverSud,
61, avenue du Président Wilson, 94230 Cachan cedex, France}


\begin{abstract}
It is shown how to model weakly dissipative free-surface
flows using the classical potential flow approach. 
The Helmholtz--Leray decomposition is applied to the linearized 3D Navier--Stokes
equations. The governing equations are treated using Fourier--Laplace transforms. We show how to express
the vortical component of the velocity only in terms of the potential and free-surface elevation. A new predominant 
nonlocal viscous term is derived in the bottom kinematic
boundary condition. The resulting formulation is simple and does not 
involve any correction procedure as in previous viscous potential flow theories 
\cite{Joseph2004}. Corresponding long wave model equations are derived.


\vskip 0.5\baselineskip

\selectlanguage{francais}
\noindent{\bf R\'esum\'e}
\vskip 0.5\baselineskip
\noindent
{\bf Ecoulements viscopotentiels avec surface libre en profondeur finie.}
Nous montrons comment modéliser les écoulements à surface libre
faiblement dissipatifs en utilisant une approche potentielle. La décomposition
de Helmholtz--Leray est appliquée aux équations de Navier--Stokes linéarisées.
Le problème est étudié au moyen de la transformée de Fourier--Laplace. Nous montrons comment exprimer la partie rotationnelle 
de la vitesse en fonction du potentiel
des vitesses et de l'élévation de la surface libre. Un nouveau terme nonlocal prépondérant
apparaît dans la condition cinématique au fond. La formulation finale est
simple et ne requiert pas de corrections ultérieures comme dans \cite{Joseph2004}.
Un modèle d'ondes longues est obtenu à partir de ces nouvelles équations.


\vskip 0.5\baselineskip
\noindent{\small{\it Mots-cl\'es~:} écoulement potentiel, écoulement avec surface libre, viscosité, dissipation, vagues,
atténuation de vague}}
\end{abstract}

\end{frontmatter}

\section*{Version française abrégée}

Récemment, Dias et al. \cite{Dias2007} ont donné une nouvelle formulation du problème des vagues en
profondeur infinie en présence de faible dissipation. L'idée principale est de partir des équations de 
Navier--Stokes linéarisées (\ref{eq:NS}) où $\nu$ représente la viscosité cinématique, $\v$ le vecteur vitesse, $\g$
l'accélération due à la gravité, $p$ la pression, $\rho$ la densité, $t$ le temps, et d'utiliser la décomposition de 
Helmholtz--Leray (\ref{eq:helmleray})
pour la vitesse. Ensuite l'on exprime la composante rotationnelle $\vpsi$ de la vitesse en fonction de la composante 
potentielle $\phi$ et de la déformation $\eta$ de la surface libre. Nous présentons ici une extension au cas
d'une couche de fluide de profondeur finie. Dans ce cas-là il faut tenir compte des frottements au fond. 

En utilisant la transformée de Fourier--Laplace, on écrit la solution générale pour $\phi$ et $\vpsi$. Ensuite l'on
utilise la condition dynamique sur la surface libre. La nullité des deux composantes tangentielles de la contrainte donne les
équations (\ref{eq:sigmazx}) et (\ref{eq:sigmazy}), qui peuvent être recombinées pour donner l'équation (\ref{eq:importantrelation}).
Cette même équation, combinée avec la condition cinématique sur la surface libre (\ref{eq:alsoimportant}), donne un lien 
entre $\eta$ et $\vpsi$ 
(\ref{eq:freesurftransform}) et un autre lien entre $\eta$ et $\phi$ (\ref{eq:roteta}). Le remplacement de la
composante rotationnelle dans la condition cinématique à la surface libre donne la première équation modifiée
(\ref{kindiss}). De même l'équilibre de la contrainte normale au niveau de la surface libre donne l'équation de
Bernoulli modifiée (\ref{berdiss}).

La seconde partie de la note consiste à incorporer une correction de type couche limite dans la condition cinématique au
fond. L'idée est d'introduire l'épaisseur de la couche limite proportionnelle à $\delta=\sqrt\nu$, la coordonnée de
couche limite $\zeta=(z+h)/\delta$, puis d'effectuer un développement du potentiel et de la vitesse rotationnelle 
en fonction du petit paramètre $\delta$. On obtient ensuite une hiérarchie de problèmes à résoudre. Ce qui nous intéresse, c'est 
la correction à apporter à la condition cinématique au fond (\ref{cinfond}). Après quelques calculs, on obtient la 
correction présentée dans l'équation (\ref{fond_dissip}). Finalement nous proposons le nouveau système d'équations 
(\ref{full1})--(\ref{full2}) pour l'étude des vagues avec faible viscosité en profondeur finie. Notons que les termes
nonlinéaires ont été ajoutés de façon heuristique. A partir de ce système,
nous pouvons dériver le système de Boussinesq faiblement dissipatif donné par les équations (\ref{bouss1})-(\ref{bouss2}).

\selectlanguage{english}

\section{Introduction}

The effects of viscosity on gravity waves have been addressed since the end
of the nineteenth century in the context of the linearized Navier--Stokes (NS) equations. 
It is well-known that Lamb \cite{Lamb1932} studied 
this question in the case of oscillatory waves on deep water. What is less
known is that Boussinesq studied this effect as well \cite{Boussinesq1895}.
In this particular case they both showed that
\begin{equation*}
  \pd{\alpha}{t} = -2\nu k^2 \alpha,
\end{equation*}
where $\alpha$ denotes the wave amplitude, $\nu$ the kinematic viscosity of the fluid
and $k$ the wavenumber of the decaying wave. This equation leads to the classical law
for viscous decay, namely $\alpha(t)=\alpha_0 e^{-2\nu k^2 t}$.

In this work we keep the features of undamped free-surface
flows while adding dissipative effects. The classical theory of
viscous potential flows \cite{Joseph2004} is based on pressure and boundary conditions
corrections due to the presence of viscous stresses. We present here a novel
approach.

Currently, potential flows with ad-hoc dissipative terms are used for example
in direct numerical simulations of weak turbulence of gravity waves
\cite{Dyachenko2003,Dyachenko2004,Zakharov2005}. There have also been several attempts
to introduce dissipative effects into long wave modelling \cite{Mei1994,Dutykh2007,CG}.

The present article is a direct continuation of the recent study \cite{Dias2007}.
In that work the authors considered two-dimensional (2D) periodic waves in infinite depth, while
in the present study we remove these two hypotheses and all the computations are done in 3D.
This point is important since the vorticity structure is more complicated in 3D.
In other words we consider a general wavetrain on the free surface of a fluid layer of finite depth. As a result 
we obtain a qualitatively different
formulation which contains a nonlocal term in the bottom kinematic condition. The inclusion
of this term is natural since it represents the correction to potential flow due to the 
presence of a boundary layer. Moreover, this term is predominant since its magnitude 
scales with $\O(\sqrt{\nu})$, while other terms in the free-surface boundary conditions 
are of order $\O(\nu)$. Other researchers have obtained nonlocal corrections but they differ from ours
\cite{KM}.

\section{Derivation}

Consider the linearized 3D incompressible NS equations
describing free-surface flows in a fluid layer of uniform depth $h$:
\begin{equation}\label{eq:NS}
  \pd{\v}{t} = -\frac{1}{\rho}\grad p + \nu\Delta\v + \g, \qquad \div\v = 0,
\end{equation}
with $\v$ the velocity vector, $p$ the pressure, $\rho$ the fluid density and $\g$ the acceleration due to gravity.
We represent $\v=(u,v,w)$ in the form of the Helmholtz--Leray decomposition:
\begin{equation}\label{eq:helmleray}
  \v = \grad\phi + \grad\times\vpsi, \qquad \vpsi = (\psi_1, \psi_2, \psi_3).
\end{equation}
After substitution of the decomposition (\ref{eq:helmleray}) into (\ref{eq:NS}), one
notices that the equations are verified provided that the
functions $\phi$ and $\vpsi$ satisfy the following equations:
\begin{equation*}
  \Delta\phi = 0, \qquad
  \phi_{t} + \frac{p-p_0}{\rho} + gz = 0, \qquad
  \pd{\vpsi}{t} = \nu\Delta\vpsi.
\end{equation*}

Next we discuss the boundary conditions. We assume that the
velocity field satisfies the conventional no-slip condition at the bottom $\left.\v\right|_{z=-h} = \0$, 
while at the free surface we have the usual kinematic condition $\eta_t = w$ and three dynamic
conditions $[\ssigma\cdot\n] = \0$, where $\ssigma$ is the stress tensor, 
$[f]$ denotes the jump of a function $f$ across the free surface, and the normal vector $\n$ equals $(0,0,1)$ due to linearization.

Using Fourier--Laplace transforms, which we denote by
$\L_\F \equiv \L\circ\F$, $f(\x,t)\stackrel{\L_\F}{\longrightarrow} \hat f(\k,s)$,
$\k=(k_x,k_y)$ we can determine the structure of the unknown functions $\phi$, $\vpsi$ in the 
transform space. We assume that all the functions involved in the present computation 
satisfy the necessary regularity requirements and have sufficient decay at infinity so that
the integral transforms can be applied.
The solution for $\phi$ is obtained from the transformed continuity equation 
$\Delta\phi=0\stackrel{\L_\F}{\longrightarrow} \fphi_{zz} - \abs{\k}^2\fphi = 0$ 
and $\vpsi$ from the corresponding transformed equation 
$\vpsi_{t} = \nu\Delta\vpsi \stackrel{\L_\F}{\longrightarrow} 
s\fvpsi = \nu\Bigl(\fvpsi_{zz} - \abs{\k}^2\fvpsi\Bigr)$:
\begin{equation*}
  \fphi = \hat\varphi_0^+(\k,s)e^{\abs{\k}z} + \hat\varphi_0^-(\k,s)e^{-\abs{\k}z}, \qquad
  \fpsi_i = \fpsi_{i0}(\k,s)\bigl(e^{\abs{m}z} + C_i(\k,s)e^{-\abs{m}z}\bigr),
\end{equation*}
where $m^2 := \abs{\k}^2 + {s}/{\nu}$ and $\hat\varphi_0^+$, $\hat\varphi_0^-$, $\fvpsi_0$, 
$\C:=(C_1,C_2,C_3)$ are
unknown functions of the transform parameters $(\k,s)$, determined by the initial and appropriate 
boundary conditions.

There are three dynamic conditions on the free surface. Let us use first those 
related to the tangential stresses (the third one will be used later), where $\mu=\rho\nu$:
\begin{equation*}
  \sigma_{xz} = \mu\Bigl(\pd{w}{x} + \pd{u}{z}\Bigr) = 0, \qquad
  \sigma_{yz} = \mu\Bigl(\pd{w}{y} + \pd{v}{z}\Bigr) = 0, \qquad \mbox{at } z=0.
\end{equation*}
Substituting decomposition (\ref{eq:helmleray}) into these two identities yields
\begin{equation}\label{eq:sigmazx}
  2\pd{^2\phi}{x\partial z} + \pd{^2\psi_2}{x^2} - \pd{^2\psi_1}{x\partial y}
  + \pd{^2\psi_3}{y\partial z} - \pd{^2\psi_2}{z^2} = 0, \qquad z=0,
\end{equation}
\begin{equation}\label{eq:sigmazy}
  2\pd{^2\phi}{y\partial z} + \pd{^2\psi_2}{x\partial y} - \pd{^2\psi_1}{y^2}
  + \pd{^2\psi_1}{z^2} - \pd{^2\psi_3}{x\partial z} = 0, \qquad z=0.
\end{equation}
The next step consists in taking the Fourier--Laplace transform to these relations.
We do not give here the explicit expressions since this operation is straightforward.
The combination 
$(-ik_x)\widehat{(\ref{eq:sigmazx})} + (-ik_y)\widehat{(\ref{eq:sigmazy})}$
gives the important relation
\begin{equation}\label{eq:importantrelation}
  ik_y\fpsi_{10}(1+C_1) - ik_x\fpsi_{20}(1+C_2) = 
  -\frac{2\abs{\k}^3}{m^2+\abs{\k}^2}(\hat\varphi_0^+ - \hat\varphi_0^-).
\end{equation}

Let us turn to the free-surface kinematic condition 
$\pd{\eta}{t} = w \equiv \pd{\phi}{z} + \pd{\psi_2}{x} - \pd{\psi_1}{y}$,
$z=0$. In transform space it becomes
\begin{equation}\label{eq:alsoimportant}
  s\feta = \abs{\k}(\hat\varphi_0^+ - \hat\varphi_0^-) + 
  ik_y\fpsi_{10}(1+C_1) - ik_x\fpsi_{20}(1+C_2).
\end{equation}
Equations (\ref{eq:importantrelation}) and (\ref{eq:alsoimportant}) can be rewritten as
\begin{eqnarray}\label{eq:freesurftransform}
  \frac{\abs{\k}(\hat\varphi_0^+ - \hat\varphi_0^-)}{\nu(m^2 + \abs{\k}^2)} & = & \feta, \\
\label{eq:roteta}
  ik_y\fpsi_{10}(1+C_1) - ik_x\fpsi_{20}(1+C_2) &= & -2\nu\abs{\k}^2\feta.
\end{eqnarray}
Using (\ref{eq:roteta}) one can replace the rotational part in the kinematic 
free-surface condition:
\begin{equation}\label{kindiss}
  \eta_{t} = \phi_{z} + \L_\F^{-1}\bigl[-2\nu\abs{\k}^2\feta\bigr] = 
  \phi_{z} + 2\nu\Delta\eta.
\end{equation}

In order to account for the presence of viscous stresses, we have to modify the
dynamic free-surface condition as well. This is done using the balance
of normal stresses at the free surface:
\begin{equation*}
  \sigma_{zz} = 0 \mbox{ at } z=0 \Rightarrow
  p-p_0 = 2\rho\nu\pd{w}{z} \equiv 2\rho\nu\Bigl(
  \pd{^2\phi}{z^2} + \pd{^2\psi_2}{x\partial z} - \pd{^2\psi_1}{y\partial z}\Bigr).
\end{equation*}
Using (\ref{eq:roteta}) one can show that 
$\pd{^2\psi_2}{x\partial z} - \pd{^2\psi_1}{y\partial z} = \O(\nu^\frac12)$, so Bernoulli's equation becomes 
\begin{equation}\label{berdiss}
  \phi_{t} + g\eta + 2\nu\phi_{zz} + \O(\nu^\frac32) = 0.
\end{equation}
Since we only consider weak dissipation ($\nu\sim 10^{-6} - 10^{-3}$ m$^2$/s),
we neglect terms of order $o(\nu)$.


The second step in our derivation consists in introducing a
boundary layer correction at the bottom. Obviously, this was not done
in the previous study \cite{Dias2007}, since the derivation dealt with the
infinite depth case. In order to include this modification, we consider a semi-infinite fluid layer as it is usually 
done in boundary layer theory. The fluid occupies the domain $z>-h$. In this derivation
we use the pure Leray decomposition of the velocity field
$\v=\grad\phi+\u$ together with the divergence-free constraint $\div\u=0$. 
Expecting that the rotational part $\u$ varies rapidly in a
distance $\delta = \sqrt{\nu}$,\footnote{Of course we should nondimensionalize all quantities in order to define
small numbers. One would find that $\delta$ is in fact equivalent to $\sqrt{Re^{-1}} \times L$, where $Re$ is the Reynolds
number and $L$ a typical length.} we introduce the boundary-layer coordinate $\zeta\equiv {(z+h)}/{\delta}$, so that 
$\u=\u(\x,\zeta) = \bigl(\u_{\x}, u_z \bigr)(\x,\zeta)$.
The solid boundary is given by $\zeta=0$, and the
potential part of the flow is not subject to this change of variables.
With the new scaling, the divergence-free condition becomes 
\begin{equation}\label{eq:divfree}
	\pd{u_z}{\zeta} + \delta\divx\u_{\x} = 0.
\end{equation}
As done in \cite{Mei1994}, we expand the unknown functions in powers of the small parameter $\delta$:
\begin{equation*}
  \phi = \phi_0(\x,z,t) + \delta\phi_1(\x,z,t) + \ldots, \quad
  \u = \q_0 (\x,\zeta,t) + \delta\q_1(\x,\zeta,t) + \ldots .
\end{equation*}
Substituting the expansion for $\u$ into (\ref{eq:divfree}) gives the 
following relations:
\begin{equation}\label{eq:divfreeAs}
  \delta^0:\quad \pd{q_{0_z}}{\zeta} = 0, \qquad
  \delta^1:\quad \pd{q_{1_z}}{\zeta} = -\divx\q_{0_{\x}},
\end{equation}
where $\q_{0_{\x}}$ denotes the first two components of the vector $\q_0$ corresponding to
the horizontal coordinates $\x$. Recall that we would like to determine the correction
to the bottom boundary condition 
\begin{equation}\label{cinfond}
	\phi_{z} = -\left.u_z\right|_{\zeta=0} = 
-\left.(q_{0_z} + \delta q_{1_z})\right|_{\zeta=0} + o(\delta).
\end{equation}
So we only need to compute $q_{0_z}$ and $q_{1_z}$ at the bottom $\zeta=0$.

Using the same asymptotic considerations as above, we can write down the following 
sequence of problems:
\begin{equation*}
  \Delta\phi_0 = 0, \quad \left.\pd{\phi_0}{z}\right|_{z = -h} = 0, \qquad\qquad
  \pd{\q_0}{t} = \pd{^2\q_0}{\zeta^2}, \quad \left.\q_0 = -\grad\phi_0\right|_{\zeta=0},
\end{equation*}
\begin{equation*}
  \Delta\phi_1 = 0, \quad \left.\pd{\phi_1}{z}\right|_{z = -h} = -q_{1_z}, \qquad\qquad
  \pd{\q_1}{t} = \pd{^2\q_1}{\zeta^2}, \quad \left.\q_1 = -\grad\phi_1\right|_{\zeta=0},
\end{equation*}
together with the radiation condition $\q\to \0$ as $\zeta\to\infty$.

This sequence of linear problems can be solved using Fourier transforms.
In Fourier space one finds immediately that 
$\fphi_0(t,z,\k) = \hat\varphi_0(t,\k)\bigl(e^{\abs{\k}z} + e^{-\abs{\k}z}\bigr)$.
Since we know $\fphi_0$, we can determine the rotational component $\hat\q_0$. 

Analytical solutions to the equation ${\partial\hat\q_{0_{\x}}}/\partial{t} = \partial{^2\hat\q_{0_{\x}}}/\partial{\zeta^2}$
are well-known. If we assume that initially the flow is potential and the boundary condition is
$\hat\q_{0_{\x}}=i\k\fphi_0(z=-h;\k)$, the solution is
\begin{equation*}
  \hat\q_{0_{\x}} = \frac{1}{2\sqrt{\pi}}\int\limits_0^t \frac{\zeta}{(t-\tau)^{\frac32}}
  e^{-\frac{\zeta^2}{4(t-\tau)}} i\k\fphi_0(\tau,z=-h,\k) \; d\tau.
\end{equation*}

Let us now integrate the second equation in (\ref{eq:divfreeAs}) from $0$ to $\infty$, using the appropriate decay at infinity:
\begin{equation*}
  \left.\hat q_{1_z}\right|_{\zeta=0} = -\int\limits_0^{\infty} i\k\cdot\hat\q_{0_{\x}}\; d\zeta = 
  \frac{1}{2\sqrt{\pi}}\int\limits_0^{\infty}
  \int\limits_0^t \frac{\zeta}{(t-\tau)^{\frac32}}
  e^{-\frac{\zeta^2}{4(t-\tau)}} \abs{\k}^2\fphi_0(\tau,z=-h,\k)\; d\tau \; d\zeta.
\end{equation*}
One can interchange integral signs and evaluate the inner integral on $\zeta$ to obtain:
  $
  \left.\hat q_{1_z}\right|_{\zeta=0} = \frac{1}{\sqrt{\pi}} 
  \int\limits_0^t\frac{\abs{\k}^2\fphi_0(\tau,z=-h,\k)}{\sqrt{t-\tau}}\; d\tau.
  $
Hence, the bottom boundary condition becomes, at order $\delta$, 
\begin{equation}\label{fond_dissip}
  \left.\pd{\phi}{z}\right|_{z=-h} = - \sqrt{\frac{\nu}{\pi}}\int\limits_0^t\frac{\F^{-1}\bigl(\abs{\k}^2\fphi_0(-h,\k)\bigr)}
  {\sqrt{t-\tau}}\; d\tau = 
  \sqrt{\frac{\nu}{\pi}}\int\limits_0^t\frac{\left.\nabla^2_{\x}\phi_0\right|_{z=-h}}{\sqrt{t-\tau}}\; d\tau
   = -\sqrt{\frac{\nu}{\pi}}\int\limits_0^t\frac{\left.\phi_{0zz}\right|_{z=-h}}{\sqrt{t-\tau}}\; d\tau.
\end{equation}
One recognizes on the right-hand side a half-order integral operator. Summarizing the developments made above and generalizing our equations by including 
nonlinear terms (this is a conjecture at this stage), we obtain a new set of viscous potential free-surface flow equations:
\begin{eqnarray} \label{full1}
  \Delta\phi &=& 0, \qquad\qquad (\x,z) \in \Omega = \R^2\times[-h,\eta] \\
  \eta_{t} + \grad\eta\cdot\grad\phi &=& \phi_{z} + 2\nu\Delta\eta, \qquad z=\eta \\
  \phi_{t} + \frac12\abs{\grad\phi}^2 + g\eta &=& -2\nu\phi_{zz}, \qquad\qquad z=\eta \\ \label{full2}
  \phi_{z} &=& -\sqrt{\frac{\nu}{\pi}}
  \int\limits_0^t\frac{\phi_{zz}}{\sqrt{t-\tau}}\; d\tau,
  \qquad z = -h.
\end{eqnarray}

Using this weakly damped potential flow formulation one can derive the following system of Boussinesq equations
with horizontal velocity $\u_h$ defined at the depth $z_\theta=-\theta h$, $ 0 \leq \theta \leq 1$: 
\begin{eqnarray}\label{bouss1}
  \eta_t + \div\left((h+\eta)\u_h\right) + h^3\left(\frac{\theta^2}{2}-\theta+\frac13\right)
  \nabla^2(\div\u_h) & = & 2\nu\Delta\eta + 
  \sqrt{\frac{\nu}{\pi}}\int\limits_0^t\frac{\div\u_h}{\sqrt{t-\tau}}\; d\tau, \\ \label{bouss2}
  {\u_h}_t + \frac12\nabla|\u_h|^2 + g\nabla\eta - h^2\theta\left(1-\frac{\theta}{2}\right)\nabla(\div{\u_h}_t)
  & = & 2\nu\Delta\u_h.
\end{eqnarray}

\section{Conclusion}

In the present paper we have shown how to express the rotational component of
the velocity field in terms of the potential part of Helmholtz--Leray decomposition.
This expression contains differential and integral operators.
Obviously, this analysis is only linear. In future work we will try to 
extend the present derivation to the nonlinear case. A long wave approximation
was derived from this new potential flow formulation.

It is interesting to note that dissipative terms of this form have been 
used to verify the theory of weak turbulence of surface gravity waves in deep water
\cite{Dyachenko2003,Dyachenko2004,Zakharov2005}. They were added without 
justification to model dissipation at small scales.
Note that a good qualitative agreement 
was obtained between the Kolmogorov spectrum
predicted by weak turbulence theory and the results of DNS.
Hence, the present work can be considered as an attempt to justify the inclusion
of these terms.

Our final remark concerns the nonlocal term in the kinematic bottom boundary
condition. This term can be also considered as a boundary layer correction
at the bottom. In modelling viscous effects this term plays the main
role, since its magnitude is $\O(\sqrt{\nu})$. Of course, the numerical 
implementation of this term is another matter.

What is the value of $\nu$
to be taken in numerical simulations? There is surprisingly little published information
of this subject. What is clear is that the molecular diffusion is too
small to model true viscous damping and one should rather consider the eddy viscosity
parameter.

\section*{Acknowledgments}

The authors are grateful to Prof. A. I. Dyachenko and Dr. D. Bouche for helpful discussions. 

\bibliographystyle{plain}
\bibliography{DutykhDiasVisco}

\end{document}